\documentclass{article}

\usepackage{arxiv}

\usepackage[utf8]{inputenc} 
\usepackage[T1]{fontenc}    
\usepackage{hyperref}       
\usepackage{url}            
\usepackage{booktabs}       
\usepackage{amsfonts}       
\usepackage{amsmath} 
\usepackage{nicefrac}       
\usepackage{microtype}      
\usepackage{lipsum}		
\usepackage{graphicx}
\usepackage{cite}
\usepackage{doi}
\usepackage{caption}

\usepackage{algorithm}
\PassOptionsToPackage{noend}{algpseudocode}
\usepackage{algpseudocode}

\algdef{SE}[DOWHILE]{Do}{doWhile}{\algorithmicdo}[1]{\algorithmicwhile\ #1}%

\title{Mastering Complex Modes: A New Method for Real-Time Modal Identification of Vibrating Systems}

\author{ \href{https://orcid.org/0000-0001-7964-5247}{\includegraphics[scale=0.06]{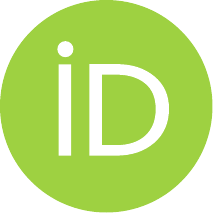}\hspace{1mm}Satyam Panda}\\
	Department of Civil Engineering\\
	Indian Institute of Technology Guwahati\\
	\texttt{panda18@iitg.ac.in} \\
	\And
	\href{https://orcid.org/0000-0002-2122-2636}{\includegraphics[scale=0.06]{orcid.pdf}\hspace{1mm}Sanghamitra Das} \\
	Department of Mechanical Engineering\\
	Indian Institute of Technology Guwahati \\
	\texttt{sanghamitra.das541@gmail.com} \\
	\And
	\href{https://orcid.org/0000-0002-4826-2232}{\includegraphics[scale=0.06]{orcid.pdf}\hspace{1mm}Basuraj Bhowmik} \\
	Department of Civil Engineering\\
	Indian Institute of Technology (BHU), Varanasi\\
	\texttt{basuraj.civ@iitbhu.ac.in} \\
	\And
	\href{https://orcid.org/0000-0001-7782-513X}{\includegraphics[scale=0.06]{orcid.pdf}\hspace{1mm}Budhaditya Hazra} \\
	Department of Civil Engineering\\
	Indian Institute of Technology Guwahati\\
	\texttt{budhaditya.hazra@iitg.ac.in} \\
}

\renewcommand{\shorttitle}{Online modal identification}

\begin{document}
	\maketitle
	
	\begin{abstract}
		A novel algorithm for real-time modal identification in linear vibrating systems with complex modes is introduced, utilizing a combination of first order eigen-perturbation and second order separation techniques. In practical settings, structures with complex modes are frequently encountered and their presence often poses a challenge in accurately estimating the source signal in real-time. The proposed methodology addresses this issue by incorporating the right angle phase shift of the response in the sensor output and updating the second order statistics of the complex response through first order eigen-perturbation. Empirical evidence of the efficacy of the technique is demonstrated through numerical case studies and validation using various numerically modeled systems, as well as a standard ASCE-SHM benchmark problem with complex modes, highlighting the capability of the proposed method to achieve precise real-time modal property identification and online source separation with a minimal number of initially required batch data.
	\end{abstract}


	\section{Introduction}
	Vibration-based condition monitoring is a diagnostic methodology utilized to detect anomalies in machinery \cite{xue2006acoustical,hazra2017fault}, aerodynamics models \cite{chauhan2010dynamic}, structures, such as buildings and bridges \cite{sadhu2017review,zhu2017thermal}, to name a few. The vibrational response of these systems under external excitation can vary depending on their structural and energy dissipation properties \cite{masciotta2022tracking}. Classical viscous damping models -- such as Rayleigh damping -- are often employed to represent the damping forces of the system for dissipative forces. Although these assumptions are accurate for simpler models, real systems often contain many different materials and are combined together in different ways to form complex spatial geometries, these assumptions are sometimes impractical and constricting. Non-classical damped linear systems can exhibit unique complex mode shapes \cite{imregun1995complex} due to non-proportional damping matrices \cite{fuellekrug2008computation}. Identification of these intricate modes is crucial for accurate characterization of the system and can be achieved through modal identification techniques, a well-researched area within vibration-based condition monitoring \cite{antoni2013study,hazra2012hybrid,brincker2015introduction,liu2009application}. However, traditional algorithms are not suited for dynamic objects, like windmills and bridges, due to their offline nature. Recent studies have introduced first order eigen perturbation (FOEP) techniques as a means to continuously update the eigenspace data and provide real-time vibration modes \cite{bhowmik2019first}. Although these methods perform well for real valued modes \cite{panda2021first}, they are not applicable to complex valued modes. In light of this, the present study proposes a novel method for real-time blind source modal identification that combines eigen perturbation and second order source separation in a recursive framework. This method enables accurate estimation of complex modes and modal response at each moment in time. 
	
	The past decade has seen widespread application of eigen perturbation theory \cite{bhowmik2019first} in the field of vibrating systems, particularly for the detection of damage instances \cite{panda2022online}, modal identification \cite{bhowmik2019first}, enhancement of system response, tuning of tune mass dampers (single and multi mode), and filtering of non-stationary signals \cite{bhowmik2022feedback}. These FOEP techniques are data-driven and exhibit reduced dependence on baseline parameters, making them a valuable tool for real-time condition monitoring. Despite their efficacy for the identification of real valued mode shapes, FOEP techniques exhibit limitations in the estimation of complex valued modes. In the context of infrastructure asset management, these methods can aid in the effective monitoring and maintenance of critical assets. Modal analysis applications continue to rely on real valued mode shapes for tuning, updating, and damage identification. However, real-time algorithms that update the eigenspace at each time instant can lead to an accumulation of errors due to the lack of consideration of the complex phasing of the system during updates. This ignorance of the complex nature of vibrational modes can result in inaccurate results for in-situ problems that require high precision. It is important to note that complex modes can exist in structures without damping -- such as those with closely spaced modes \cite{mcneill2008framework}, subjected to aerodynamic loads \cite{d2013analyzing}, or characterized by asymmetry \cite{goel2000seismic}. Ignoring these complex modes in modal analysis can negatively impact the accuracy of results.

	The traditional real-time FOEP algorithms update the eigenspace of the system through numerical simulation or sensor-acquired data. The eigenvector is then checked for permutation ambiguity to determine the modes of the system. However, many systems possess inherently complex modes that are overlooked by these algorithms. To overcome this limitation, the proposed algorithm involves taking two simultaneous samples and performing a Hilbert transform to obtain a 90$^\circ$ phase shift, resulting in a complex response. The FOEP algorithm is then used to update the complex eigenspace with the updated eigenspace serving as the basis for estimating the whitening matrix and updating the delayed covariance matrix. This novel approach offers improved accuracy in characterizing complex modes, which has significant commercial and sectorial benefits in the field of civil engineering, including improved infrastructure asset management, enhanced vibration-based damage identification, and advanced tuning and filtering techniques. The proposed real-time algorithm employs the principles of eigen perturbation and second order source separation to accurately estimate the complex-valued modes of ever-changing systems. This is accomplished through the simultaneous acquisition of two samples, the calculation of the Hilbert transform, and the updating of the eigenspace and delayed covariance matrix using FOEP techniques. The resulting complex de-mixing matrix separates the source signals, enabling the estimation of complex modes in real-time. The development of this algorithm provides significant benefits for the commercial and industrial sector, particularly in the field of civil engineering, by enabling the accurate and efficient assessment of infrastructure assets.
	
	The proposed framework in this study constitutes a major advancement in the field of real-time condition monitoring for civil and structural engineering applications. Specifically, it offers the following key contributions:
	
	\begin{enumerate}
		\item Development of a mathematically consistent algorithm for the real-time estimation of complex modes in dynamic systems, which has been an understudied area in the field of structural health monitoring and damage detection.
		\item Incorporation of second-order delayed statistics updates into traditional FOEP techniques to account for the constantly changing complex phasing in systems with dynamic modes.
		\item The proposed framework is applicable for both real and complex modal identification in a variety of simulations and practical problems, providing a more comprehensive solution for the blind source problems faced in real-life structural health monitoring and damage detection applications.
	\end{enumerate}
	
	Overall, this study represents a significant step towards the accurate identification of complex modes in real-time, which is a crucial aspect in the development of effective structural health monitoring and damage detection systems.

	The paper's structure comprises of the following segments: Section 2 delineates a succinct exposition of the complex state theory within the purview of structural dynamics and the conventional batch Blind Source Separation (BSS) technique, namely the Second Order Blind Identification (SOBI), utilized for the calculation of complex modes. Section 3 propounds a mathematically rigorous formulation of the proposed methodology for determining complex modes of general dynamic systems, incorporating an algorithm and flowchart. Section 4 presents numerical simulations on selected categories of linear systems to exemplify the application of the proposed methodology in both simulation and real-world situations, validating its efficacy against the Structural Health Monitoring (SHM) benchmark structure outlined by the International Association for Structural Control and Monitoring - American Society of Civil Engineers (IASC-ASCE). Lastly, Section 5 summarizes the work and provides insightful observations.

	\section{Background}
	This section explains the necessity of identification of the complex modes in structural systems with a concise summary of blind modal identification techniques.
	\subsection{Complex modes in structural dynamics}
	The concept of complex modes in the light of structural dynamics can be understood by considering the generic dynamical system of the form: 
	where, mass, stiffness and damping matrices are represented by \textbf{M}, \textbf{C} and \textbf{K} with X as the displacement vector and $\textbf{F}(t)$ as the input excitation. Since the matrices \textbf{M}, \textbf{C} and \textbf{K} are diagonalized in the modal coordinate the modal responses (${\bf{Q}} = {[{q_1},{q_2}, \ldots {q_i} \ldots {q_n}]^T}$) of the Eqn. Eq. \ref{basic} can be written as:
	\begin{equation}\label{basic_modal}
		{{\ddot q}_i}(t) + 2{\zeta _i}{\omega _{n,i}}{{\dot q}_i}(t) + {\omega ^2}_{n,i}{q_i}(t) = {\frac{1}{m_i}}{f_i}(t)
	\end{equation}
	where, ${f_i}(t)={v _i}^T{\bf{F}}(t) $ is the modal force corresponding to the ${v _i}$ mode. $\zeta _i$, ${\omega ^2}_{n,i}$  and $m_i$ are the modal damping, modal frequency and modal mass for $i^{th}$ mode. The solution of the Eq. \ref{basic_modal} is obtained by using Duhamel Integral: ${q_i}(t) = \int\limits_0^\infty  {{f_i}(\tau ){h_i}(t - \tau )} {\kern 1pt} d\tau $, where ${h_i}(t - \tau ) = \frac{1}{{{m_i}{\omega _{d,i}}}}{e^{ - {\zeta _i}{\omega _{n,i}}(t - \tau )}}\sin {\kern 1pt} {\kern 1pt} {\omega _{d,i}}(t - \tau )$ and ${\omega ^2}_{d,i}$ is the damped modal frequency. The solution matrix ${\bf{Q}}$ in the vectorial form can be expressed as (using convolution property):
	
	\begin{equation}\label{basic}
		{\bf{M\ddot X}}(t) + {\bf{C\dot X}}(t) + {\bf{KX}}(t) = {\bf{F}}(t)
	\end{equation}
	
	\begin{equation}
			{{\bf{Q}}_{i \times 1}}(t) = \int\limits_0^\infty  {\underbrace {{{\left[ {\begin{array}{*{20}{c}}
									{{h_1}(t - \tau )}&0& \cdots &0\\
									0&{{h_2}(t - \tau )}& \cdots &0\\
									\vdots & \vdots & \ddots & \vdots \\
									0&0& \cdots &{{h_i}(t - \tau )}
							\end{array}} \right]}_{i \times i}}}_{{{\bf{H}}^Q}(t - \tau )}} \underbrace {{{\left\{ {\begin{array}{*{20}{c}}
								{{f_1}(\tau )}\\
								{{f_2}(\tau )}\\
								\vdots \\
								{{f_i}(\tau )}
						\end{array}} \right\}}_{i \times 1}}}_{{{{F}}^Q}(\tau )}d\tau 
	\end{equation}
	The solution of Eq. (\ref{basic}) can be written as ${\bf{X}} = {\bf{VQ}}$, where \textbf{X} is the measurement matrix of size  $m\times N$ and \textbf{Q} is the corresponding modal response of size $m\times N$ with $m$ as number of degrees of freedom and $N$ as the sampling size. \textbf{V} is a orthogonal transformation matrix of size $m \times m$ yielding mode matrix such that the mode shape matrix are orthogonal to each other with respect to the matrix \textbf{M}. In general, mode shape matrices of structural dynamical systems, whether with constant system parameters or with varying mass and stiffness properties, are often approximated as proportional modes. However, a complex modal solution is necessary when addressing system modifications with respect to factors such as damping or non-linearity. If proportional mode approximations are used, errors can accumulate due to the complex phasing resulting from the modified system, which the proportional approximation overlooks.
	
	Complex modes can arise in non-proportionally damped systems either for a narrow range of system parameters or for systems with closely spaced modes. Complex modes can also occur in aeroelastic systems under non-linear aerodynamic loads. Furthermore, the presence of analysis and measurement errors can also lead to complex modes \cite{imregun1995complex}. Complex eigenvectors can also result from an asymmetric velocity-dependent system matrix, even if there is no damping and only a gyroscopic term present in the ${\bf{C}}$ matrix. This complexity can also arise in axisymmetric structures, where modes are often identical-eigenvalue pairs and hence, have a degree of degeneracy. It is crucial for practitioners dealing with real-time modal identification problems to utilize algorithms that can handle both proportional and complex modes to avoid incorrect modal identification. This necessitates the effective use of Blind Source Separation (BSS) algorithms with specific capabilities for real-time mechanism and complex modal identification. To address these problems, a succinct understanding of the BSS approach and its application to structural dynamics is required, which is discussed in the following section.
	
	\subsection{Blind source separation (BSS)}
	Blind Source Separation (BSS) refers to the process of extracting source signals from output sensor signals without any prior knowledge of the system. In BSS, only the mixed output signals are available for separation in order to retrieve the source signals. Numerous studies have been conducted to solve modal identification problems using BSS, which can act as a useful inverse methodology for retrieving the source signals of an unknown Multiple-Input/Multiple-Output (MIMO) mixing system. The input signals are estimated based on the output sensor signals. Several BSS techniques for estimating the source signals include Independent Component Analysis (ICA) and its extensions (Topographic ICA, Multidimensional ICA, Kernel ICA, Tree-dependent Component Analysis, Subband Decomposition -ICA), Sparse Component Analysis (SCA), Sparse PCA (SPCA), Non-negative Matrix Factorization (NMF), Smooth Component Analysis (SmoCA), Parallel Factor Analysis (PARAFAC), Time-Frequency Component Analyzer (TFCA), and Multichannel Blind Deconvolution (MBD) \cite{choi2005blind}.
	BSS finds wide applications in many areas, such as array signal processing, seismic signal processing, and blind equalization \cite{cao1996General}. It assumes that the output data is a linear combination of the modal coordinates, which are estimated from the output signals only. This makes BSS a useful technique in the estimation of modal responses from ambient data. By comparing the superposition of vibration modes with the BSS model's specification, it is possible to understand the physical significance of the sources in system identification problems. The mode shapes, natural frequencies, and damping factors of a vibrating structure can be estimated using BSS, providing a numerical solution for the deflection patterns of vibration when the system vibrates at one of its natural frequencies.  BSS has various applications in the field of signal processing and engineering, including array signal processing, seismic signal processing, and blind equalization. By assuming the output data is a linear combination of modal coordinates and only using the output signals for estimation, linear mixing models can be used to estimate modal responses from ambient data. These models can then be used to determine the natural frequencies, damping factors, and mode shapes of a vibrating structure, which provide insight into its physical behavior during vibration. BSS plays an important role in system identification problems by facilitating the comprehension of the physical importance of sources. 
	
	\subsection{Second Order Blind Identification (SOBI)}
	For system identification problems, the popular BSS technique known as Second Order Blind Identification (SOBI) is frequently used \cite {belouchrani1997blind}.SOBI has a wide variety of applications in modal identification of vibrating structures \cite {mcneill2008framework}. It has proven to be a viable substitute to produce accurate results for output-only modal analysis of vibrating structures \cite {rainieri2014perspectives}. Independent component analysis (ICA) is found to give better results for identification of mode shapes as compared to Principal component analysis (PCA). The reason behind this is the poor estimation performance and high computation time at higher order statistics and damping ratios greater than 1 percent \cite {belouchrani1997blind}. A suitable alternative to estimate the mode shapes for second order statistics is the second order blind identification (SOBI) which assumes uncorrelated sources with variable spectral contents\cite{kerschen2007physical}. SOBI estimates real valued mode shapes with real values mixing matrix. It uses the time information contained in the signals in contrary to considering them as random variables in ICA. SOBI has also shown to be a promising technique for time series dimension reduction.
	\newline SOBI is based on the premise of simultaneous diagonalization of two covariance matrices ${\bf \hat R}_{\bf Y} \left( 0 \right)$ and ${\bf \hat R}_{\bf Y} \left( p \right)$, defined by the relations as under:
	
	\begin{equation}
		\left. \begin{array}{l}
			{\bf R}_{ Y} \left( p \right) = E\left\{ {{ Y}\left( k \right){ s}^{ T} \left( {k - p} \right)} \right\} \\
			{\bf R}_{ Y} \left( 0 \right) = E\left\{ {{ Y}\left( k \right){ Y}^{ T} \left( k \right)} \right\} = A{\bf R}_{ Y} \left( 0 \right)A^T  \\
			{\bf R}_{ Y} \left( p \right) = E\left\{ {{ Y}\left( k \right){ Y}^{ T} \left( {k - p} \right)} \right\} \\
		\end{array} \right\}\label{covmat}
	\end{equation}
	for some non-zero time-lag $p$. The simultaneous diagonalization is performed in three basic steps: whitening, orthogonalization, unitary transformation. A framework for second order blind identification method can be summarized as follows: 
	\begin{enumerate}
		\item Obtain output dataset, assign to ${{Y}_{0}}\,(t)$. Compute cross correlation functions from the data, and assign to ${{Y}_{0}}\,(t)$.
		\begin{equation} {{R}_{Y\,}}(0)\,=\,\frac{1}{N-1}\,{{Y}^{T}}
		\end{equation}
		
		\item Whiten the data to obtain whitened data, $\bar{Z}$. Retain the whitening matrix and its inverse. 
		Whitening is a linear transformation in which ${\bf \hat R}_{\bf Y} \left( 0 \right) = \left( {{\raise0.7ex\hbox{$1$} \!\mathord{\left/
					{\vphantom {1 N}}\right.\kern-\nulldelimiterspace}
				\!\lower0.7ex\hbox{$N$}}} \right)\left( {\sum\limits_{K = 1}^N {\textbf{Y}\left( k \right)\textbf{Y}^T \left( k \right)} } \right)$ is first diagonalized using singular value decomposition that is accomplished as ${\bf \hat R}_{\bf Y} \left( 0 \right) = {\bf V}_{\bf Y} {\bf \Lambda }_{\bf Y} {\bf V}_{\bf Y}^{\bf T}$.	 The whitened signal is given as:
		\begin{equation}
			{Y}\left( k \right) = {\bf W}Y\left( k \right) = {\bf \Lambda }_{Y}^{ - \frac{1}{2}} {\bf V}_{Y}^{\bf T} {Y}\left( k \right)
		\end{equation}
		where ${\bf W}$ is the whitening matrix.
		
		\item Apply joint approximate diagonilization (JAD) to the whitened data to obtain the joint diagonalizer.
		Orthogonal transformation is applied to diagonalize the matrix ${\bf \hat R}_{Z} \left( p \right)$ The Eigen value decomposition of  ${\bf \hat R}_{Z} \left( p \right)$ has the form 
		\begin{equation} {\bf \hat R}_{Z} \left( p \right) = {\bf U}_{Z} {\bf \Sigma}_{Z} {\bf U}_{Z}^{T}. 
		\end{equation}
		
		\item Compute the de-mixing matrix, $\hat{\bf A}$, and the mixing matrix, ${\bf A}$.
		\begin{equation} {\bf \hat A} = \,{\bf U}_{Z}^{\dagger}\,{\bf W} \end{equation}
		If the diagonal matrix ${\bf \Sigma}_{Z}$ has distinct eigen values then the mixing matrix can be estimated uniquely by the following equation:
		\begin{equation} 
			{\bf A}= \textbf{W}^{+} {\bf U}_{Z}
		\end{equation}
		where, $(.)^+$ is the Moore-Penrose pseudoinverse of a matrix. 
		\item Compute the modal response.
		$X\,=\,{{\bf A}^{T}}\,Y$
	\end{enumerate}
	
	\subsection{General eigen perturbation theory}
	Recently, advancements have been made in the field of real-time estimation of vibrational modes through the implementation of eigen perturbation techniques. The focus of these algorithms lies in the recursive update of the eigenspace, providing an alternative method for updating the full rank data covariance matrix. The proposed approach utilizes the framework of First-Order Eigen Perturbation (FOEP) technique for the recurrent update of the complex eigenspace, thereby updating the delayed whitening covariance matrix. As a preface to the mathematical intricacies of the proposed algorithm, it is imperative to have a comprehensive understanding of the generalized higher order eigen perturbation theory and the derivation of FOEP from it. In the context of a vibrating system, the symmetric eigen decomposition can be mathematically represented as:
	\begin{equation}\label{evp}
		{\bf{R}}_Y {\bf{V}} = {\bf{\Lambda }} {\bf{V}}
	\end{equation}
	where, ${{\bf{R}}_X} \in {\mathbb{R}^{n \times n}}$, and ${{\bf{\Lambda }}}$, ${{\bf{V}}}$ are eigenvalues and eigenvectors matrices, respectively. As the eigenvector matrix is symmetric and eigenvalue matrix is diagonal, the following orthogonality conditions will hold
	\begin{equation}\label{ortho} 
		\begin{array}{l}
			{V}_{j}^T{{V}_{i}} = {{\delta _{ij}}}, \quad
			V_{j}^T{\bf{R}}_Y{V_{i}} = {\Lambda _{i}}{\delta _{ij}} \quad \forall \quad i,j = 1,2,3, \ldots ,n
		\end{array}
	\end{equation}
	where ${\delta _{ij}}$ is the Kronecker delta. For a generalized $n^{th}$ order perturbation, Eq. (\ref{evp}) can be written as a linear combination of $n$ perturbation terms added to the original matrix, given by the expression: 
	
	\begin{equation}
		\begin{array}{l}
			{{\bf{R}}_Y} = \left( {\bf{V}} + \sum_{i=1}^{n} \delta^i {\bf{V}} \right)^T	\left( {{\lambda} + \sum_{i=1}^{n} \delta^i \lambda} \right)\left( {{{\bf{V}}} + \sum_{i=1}^{n} \delta^i {\bf{V}}} \right) 
		\end{array}
	\end{equation}
	This generalized eigen perturbation theory provides a method for updating the data covariance matrix without requiring repeated eigen decomposition of the data covariance matrix at each time step. However, the higher the perturbation order involved in the expansion, the more accurate the algorithm becomes, but also the more mathematically complex and computationally demanding. Hence, real-time system monitoring algorithms generally adopt either First \cite{bhowmik2019first} or Second Order perturbation terms \cite{mucchielli2020higher}. Nevertheless, recent studies \cite{panda2021first,panda2022online,bhowmik2022feedback} have incorporated error correction mechanisms to mitigate the impact of neglected higher-order perturbation terms and enhance the convergence and damage detection capability of traditional FOEP techniques. However, in this work, the focus is on the recursive whitening of signals using the eigenspace update of complex data, therefore the use of higher-order perturbation or error feedback mechanisms would increase computational complexity without significantly improving the robustness of the algorithm. The proposed framework combines the First Order Eigen Perturbation (FOEP) technique with the concept of Second Order Blind Source Separation (SOBSS) in a recursive manner, providing a mathematically consistent approach to the estimation of complex modes, source signal extraction, and will be discussed in detail in the following section.
	
	\section{Recursive modal identification in dynamical systems}
	The basic principle of the recursive modal identification of the systems with complex mode shapes primarily premise on three primary segments operating simultaneously in a recursive framework: \textit{Firstly} whitening of the complex response using RPCA to obtained the whitening matrix, its inverse and whitened data for each sample, \textit{Secondly} updating the lagged covariance matrix and \textit{Finally} estimation of unitary matrix from joint diagonalization of the updated lagged covariance matrix and in turn estimating the mixing matrix. These process require an initialization of some parameters which are obtained using the batch algorithms for a few initial samples (ideally 100-500samples). 
	
	The incorporation of the phase shifted data obtained through the Hilbert transformation enhances the accuracy of the real-time modal identification algorithm in handling non-proportional damping, non-linearity, gyroscopic effect, and closely placed modes, among others. The real-time feeding of the sensor data and phase shifted data to the algorithm forms a complex response, thereby allowing for more accurate identification of the modal response in an online fashion. Thus the proposed algorithm supplements the recorded response $Y(t)$ with the phase shifted data $Y_{90}(t)$ obtained through employing the Hilbert transformation (HT) and taking its imaginary value as 90$^\circ$ phase shifted response. The key point is that the response is fed to the algorithm in real time to obtain the phase shifted data and is supplemented to the sensor data for formation of the complex response as, 
	\begin{equation}\label{phshift}
		{\bar{Y}}(t) = Y(t) + \text{i} \ Y_{90}(t)
	\end{equation} 
	This leads to the problem of finding the complex valued modal response through recursive BSS techniques and can be written in the form of complex modal response as,
	\begin{equation}\label{phshift2}
		\left\{Y_k + \text{i} \ Y_{90,k}\right\} = \left[{\bf{A}}_{0,k} + \text{i} \ {\bf{A}}_{90,k}\right] \left\{X_k + \text{i} \ X_{90,k}\right\}
	\end{equation}
	After obtaining the complex response, the objective shifts towards the estimation of the whitening matrix, where, the mathematically consistent formulation of RPCA for the complex data is necessary. Towards understanding the mathematical formulation of RPCA, consider the co-variance matrix of the form: ${{\bf{C}}_Y} = \frac{1}{N}{\bf{Y}}{{\bf{Y}}^{{T}}}$. For any multivariate data set the recursive estimation of the data co-variance matrix (${{\bf{C}}_k}$) at time instant $k$ can be expressed in terms of the data vector at $k^{th}$ instant i.e. ${Y_k}$ as:
	\begin{equation}\label{rechank}
		{{\bf{C}}_k} = \frac{{k - 1}}{k}{{\bf{C}}_{k-1}} + \frac{1}{k}\left\{Y_k + \text{i} \ Y_{90,k}\right\}\left\{Y_k + \text{i} \ Y_{90,k}\right\}^T
	\end{equation}
	where ${{\bf{C}}_{k - 1}}$ is the co-variance estimate at $(k-1)^{th}$ instant. The primary aim of RPCA is to estimate the update of co-variance matrix at each time instant without actually performing EVD on the block co-variance matrices thereby reducing the time and memory complexity. First order eigen perturbation (FOEP) technique facilitates the estimation of eigenspace without actually performing the EVD recursively. The individual data co-variance estimates at $k^{th}$ instant can be expressed by its EVD as, ${{\bf{C}}_k} = {{\bf{V}}_k}{\bf{\Sigma }}_k{{\bf{V}}_k}^T$ where ${\bf{V}}_k$ and ${\bf{\Sigma }}_k$ represents the eigenvector and eigenvalue matrices at $k^{th}$ instant. For non-stationary processes the drift in mean level can be accommodated in the estimate of co-variance update by using mean shift in the data samples as follows:
	\begin{align}\label{maineqnfop}
		{{\bf{V}}_k}{\bf{\Sigma }}_k{{\bf{V}}_k}^T = \frac{{(k - 1)}}{k}{{\bf{V}}_{k - 1}}{\bf{\Sigma }}_{k - 1}{{\bf{V}}_{k - 1}}^T + \frac{1}{k}\left[ {\left\{Y_k + \text{i} \ Y_{90,k}\right\} - {\mu _k}} \right]{\left[ {\left\{Y_k + \text{i} \ Y_{90,k}\right\} - {\mu _k}} \right]^T}
	\end{align}
	where, the recursive mean at $k^{th}$ instant is estimated as: ${\mu _k} = \frac{{k - 1}}{k}{\mu _{k - 1}} + \frac{1}{k}\left\{Y_k + \text{i} \ Y_{90,k}\right\}$. However, for structural systems the response data vectors are generally evolve as a zero mean processes enabling the Eq. \ref{maineqnfop} to be written without any mean shift. Further simplifying the Eq. \ref{maineqnfop} with the relation $\left\{Y_k + \text{i} \ Y_{90,k}\right\} = {{\bf{V}}_{k-1}}{P_k}$ one gets:
	\begin{equation}\label{fopdominant}
		{{\bf{V}}_k}k{\bf{\Sigma }}_k{{\bf{V}}_k}^T = {{\bf{V}}_{k - 1}}\{ (k - 1){\bf{\Sigma }}_{k - 1} + {P_k}{P_k}^T\} {{\bf{V}}_{k - 1}}^T	
	\end{equation}	
	Under finitely large sample-size $k$ the term $\{ (k - 1){\bf{ \Sigma }}_{k - 1} + {P_k}{P_k}^T\}$ exhibits a diagonally dominant characteristics since ${\bf{\Sigma }}_{k - 1}$ is diagonally dominant. Towards obtaining a mapping between both the sides of Eq. \ref{fopdominant},  Gershgorin's theorem ensures the EVD of the term to be of the form ${\bf{\Psi }}_k{{{\bf{\Omega }}_k}}{\left( {{\bf{\Psi }}_k} \right)^T}$, where ${{\bf{\Psi }}_k}= \left({\bf I} + \delta {\bf V} \right)$ and ${{\bf{\Omega }}_k} = \left({{\bf{\Sigma }}_{k-1}} + \delta {\bf \Sigma} \right) $ are the orthonormal eigenvector and eigenvalue perturbation matrices, respectively, at $k^{th}$ instant. Upon substitution and simplification the Eq. \ref{fopdominant} can be rewritten as follows:
	\begin{equation}\label{subsevd1}
		{{\bf{V}}_k}k{\bf{\Sigma }}_k{{\bf{V}}_k}^T = {{\bf{V}}_{k-1}}{\bf{\Psi }}_k{ {{\bf{\Omega }}_k}}{\left( {{\bf{\Psi }}_k} \right)^T}{{\bf{V}}_{k - 1}}^T
	\end{equation}
	On close observation of the Eq. \ref{subsevd1}, final updates for the eigenvector and eigenvalues can be identified as:
	\begin{equation}\label{rmssa_last}
		{\bf{V}}_k = {{\bf{V}}_{k-1}}{\bf{\Psi }}_k \quad \& \quad {\bf{\Sigma }}_k = \frac{1}{k}{{\bf{\Omega }}_k}
	\end{equation}
	A drawback for using FOEP technique to estimate the eigenspace at each instant is associated with permutation ambiguity i.e. the recursive eigenvectors and eigenvalues are not arranged in same sequence as the theoretical values. However the shortcoming can be overcome by rearranging the eigenvalues in decreasing order, and, correspondingly the eigenvectors. The complex whitening data at a particular time instant can be extracted as follows:
	\begin{equation}\label{recmain}
		\left\{Z_k + \text{i} \ Z_{90,k}\right\} = {{\bf{\Sigma}}_k}^{-1/2}{{\bf{V}}_k}^T\left\{Y_k + \text{i} \ Y_{90,k}\right\} = {{\bf{W}}_k}^T\left\{Y_k + \text{i} \ Y_{90,k}\right\}
	\end{equation}
	Here, $Z_k$ is the whitening data and $Z_{90,k}$ is its 90$^\circ$ phase shift of the whitening data, and, ${{\bf{W}}_k}^T$ is the whitening matrix. This accomplishes the second objective of the proposed algorithm and the next step is to update the lagged covariance matrix formed through the whitened data. For a series of lags ($\tau = 1,\ldots,\mathcal{T}$), the augmented lagged whitening covariance matrix is formed as follows,
	\begin{equation}\label{Rlag}
		\left[ {\begin{array}{*{20}{c}}
				{{\bf{R}}_{k,k-1}}\\
				{{\bf{R}}_{k,k-2}}\\
				\vdots \\
				{{\bf{R}}_{k,k-\mathcal{T}}}
		\end{array}} \right] = \frac{{k - 1}}{k}\left[ {\begin{array}{*{20}{c}}
				{{\bf{R}}_{(k-1),(k-1)-1}}\\
				{{\bf{R}}_{(k-1),(k-1)-2}}\\
				\vdots \\
				{{\bf{R}}_{(k-1),(k-1)-\mathcal{T}}}
		\end{array}} \right] + \frac{1}{k}\left[ {\begin{array}{*{20}{c}}
				\left\{Z_k + \text{i} \ Z_{90,k}\right\}\left\{Z_{k-1} + \text{i} \ Z_{90,k-1}\right\}^T\\
				\left\{Z_k + \text{i} \ Z_{90,k}\right\}\left\{Z_{k-2} + \text{i} \ Z_{90,k-2}\right\}^T\\
				\vdots \\
				\left\{Z_k + \text{i} \ Z_{90,k}\right\}\left\{Z_{k-\mathcal{T}} + \text{i} \ Z_{90,k-\mathcal{T}}\right\}^T
		\end{array}} \right]
	\end{equation}
	where, ${\bf{R}}_{k,k-\tau}$ is the lagged whitening covariance matrix with lag of $\tau$. After obtaining augmented lagged whitening covariance matrix the problem of modal identification reduces to finding a best suitable candidate for unitarily diagonalizing $\left[{\bf{R}}_{k,k-1} \ {\bf{R}}_{k,k-2} \ \ldots \ {\bf{R}}_{k,k-\mathcal{T}}\right]$ at each instance of time. Denoting $\mathbf{D}^\tau = \mathbf{U}_k^T\left[{\bf{R}}_{k,k-\tau}\right]\mathbf{U}_k$, the problem of finding the minimum of the performance index $J$ can be mathematically written as,
	\begin{equation}
		J({\bf U},\tau)\mathop {\rm  = }
		\sum\limits_{\mathcal{T}} {\sum \limits_{1 \le i \ne j \le m} {\left| {D_{ij}^\tau } \right|^2 } }
		\label{Joint Diagonality_mcc}
	\end{equation}
	Then, for minimum $J$ over fixed $h$ iterations the unitary matrix $\mathbf{U}$ is said to be an approximate joint diagonalizing the augmented lagged whitening covariance matrix. Having obtained the whitening and unitary matrix, the mixing matrix is obtained using the following relation
	\begin{equation}\label{mixmat}
		\left[{\bf{A}} + \text{i} \ {\bf{A}}_{90}\right] = \mathbf{U}^\dagger \mathbf{W} 
	\end{equation}
	The mixing matrix is then operated on the obtained sensor response $Y_k$ to estimate the complex modal response as,
	\begin{equation}\label{modalresp}
		\left\{\left(X\right)_k + \text{i} \ X_{90,k}(t)\right\} =\left[{\bf{A}}_{0,k} + \text{i} \ {\bf{A}}_{90,k}\right]^\dagger Y_k
	\end{equation}
	Thereafter, both the complex identified mode shape matrix and modal response are normalized to obtain the real valued modal mixing matrix and responses which are independent to each other.
	\begin{equation}\label{rescaling1}
		\begin{array}{ll}
			\left({\bf{A}}_{ij}\right)_k &= \frac{1}{k}\|X_k + \text{i} \ X_{90,k}\|_i \ sgn\left(\left({\bf{A}}_{ij}\right)_{0,k}\right)  \left\|\left({\bf{A}}_{ij}\right)_{0,k} + \text{i} \ \left({\bf{A}}_{ij}\right)_{90,k}\right\| \\
			X_k	&= \frac{1}{k}\|X_k + \text{i} \ X_{90,k}\| \cdot X_k
		\end{array}
	\end{equation}
	
	\begin{figure}
		\centering
		\includegraphics[width=0.75\textwidth]{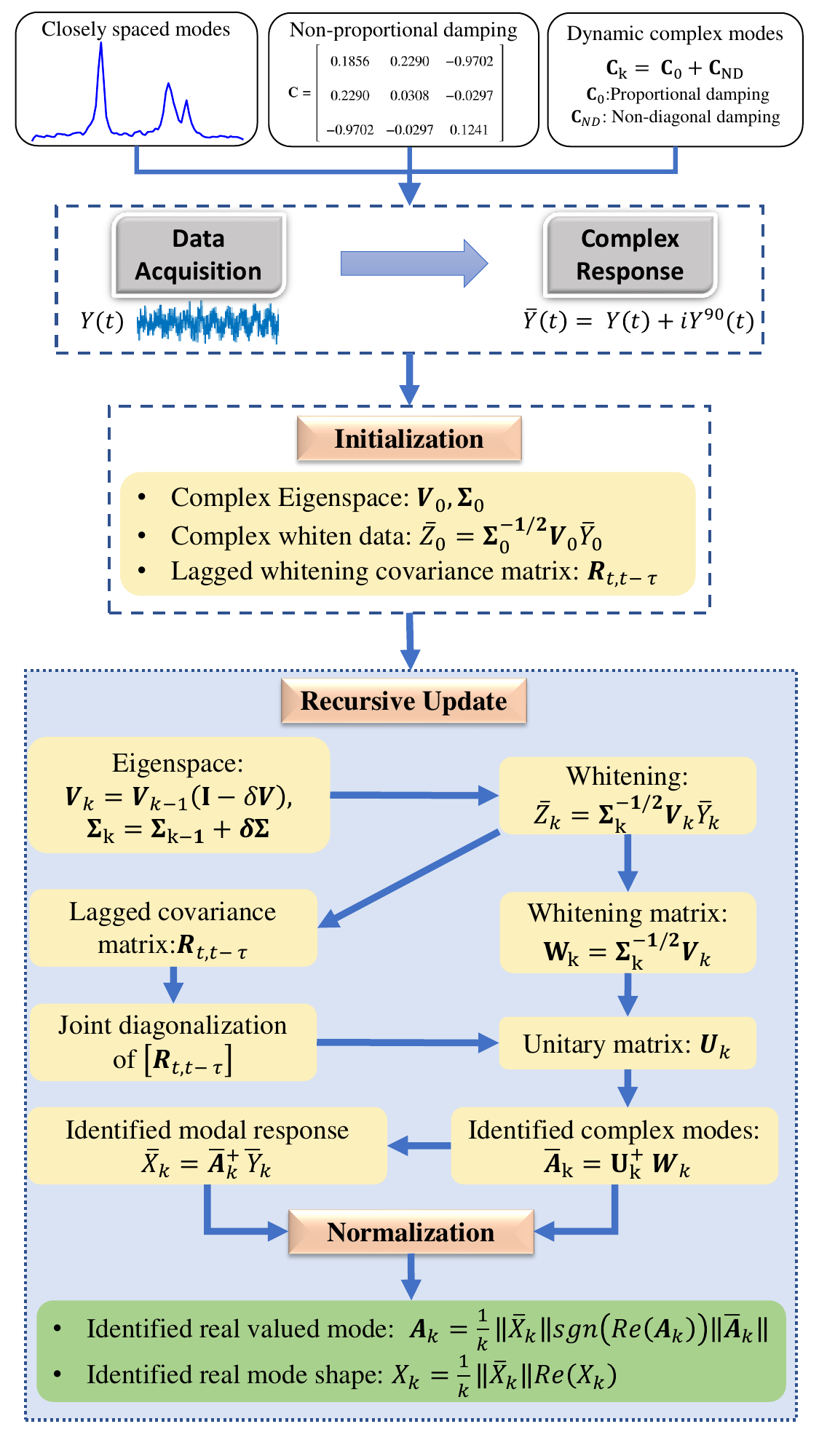}
		\caption{A framework for the proposed technique}
		\label{flow}
	\end{figure}
	
	\begin{algorithm}
		\caption{Algorithm for the proposed method}\label{euclid}
		\small
		\begin{algorithmic}
			\Require{${\bf{X}} = [X_1,X_2,...,X_L]^T \quad \forall \quad X \in {\mathbb{R}^n}$} \Comment{Response vector of system}
			\State Appending 90$^\circ$ phase shifted response \Comment{Complex response, Eq. (\ref{phshift})}
			\State \textbf{Initialization}: complex eigenspace, whiten data and its lagged covariance matrix
			\For{$k = 1$ : $N_{iteration}$}\Comment{Iteration for sample length}
			\State define memory depth parameter $\lambda$
			\State  Define ${P_k} = {{\bf{V}}_{k-1}^T \left\{Y_k + \text{i} \ Y_{90,k}(t)\right\}} $
			\State Estimation of the perturbation matrices ${\bf{\Psi}}_k$ and ${\bf{\Omega}}_k$
			\State Update of eigenvector and eigenvalue matrices \Comment{Eigenspace update, Eq. (\ref{rmssa_last})}
			\State Estimation of whitened data $	\left\{Z_k + \text{i} \ Z_{90,k}(t)\right\}$ and whitening matrix ${\bf{W}}_k$ \Comment{Eq. \ref{recmain}}
			\State Update of lagged whitened covariance matrix ${\bf{R}}_{k,k-\tau}$ \Comment{Eq. (\ref{Rlag})}
			\State Joint approx diagonalization on ${\bf{R}}_{k,k-\tau}$: $\mathbf{D}^\tau = \mathbf{U}_k^T\left[{\bf{R}}_{k,k-\tau}\right]\mathbf{U}_k$
			\State Estimation of complex mixing matrix using whitening matrix ${\bf{W}}_k$ and unitary matrix ${\bf{U}}_k$ \Comment{Eq. (\ref{mixmat})}
			\State Estimation of complex modal response  \Comment{EVD of Covariance, Eq. (\ref{modalresp})}
			\State Estimation of real valued modal matrix and response \Comment{Normalization, Eq. (\ref{modalresp})}
			\EndFor\label{euclidendwhile}
			\Ensure{Output: Normalized real valued modes (${\bf A}_k$) and  Modal response (${X}_k$)}
		\end{algorithmic}
	\end{algorithm}
	
	\begin{figure}
		\centering
		\includegraphics[width=0.25\textwidth]{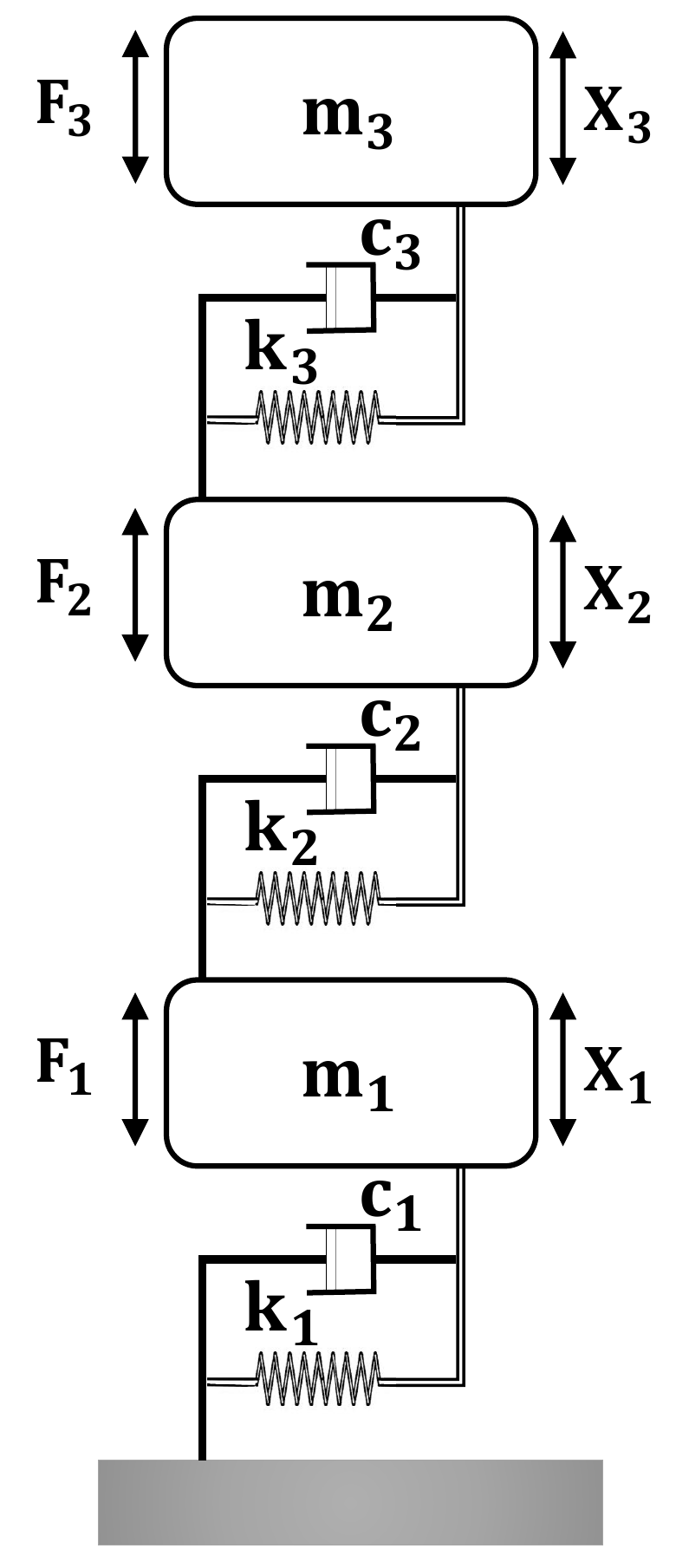}
		\caption{Representation of 3DOF linear lumped mass model}
		\label{3DOF}
	\end{figure}
	
	\section{Case studies for systems with complex mode shapes}
	The proposed algorithm's efficacy in identifying the modal response of structural systems with complex modes is demonstrated through numerical case studies on linear systems with closely spaced modes and non-proportional damping. The algorithm's performance is evaluated on three structures: (i) a 3-DOF linear structural system with closely spaced modes, (ii) a 3-DOF linear system with non-proportional damping, and (iii) an IASC-ASCE SHM benchmark structure with complex modes. The numerical simulation results and modal identification outcomes for these structures are presented and discussed in detail.
	\begin{figure}[htbp]
		\centering
		\includegraphics[width=0.8\textwidth]{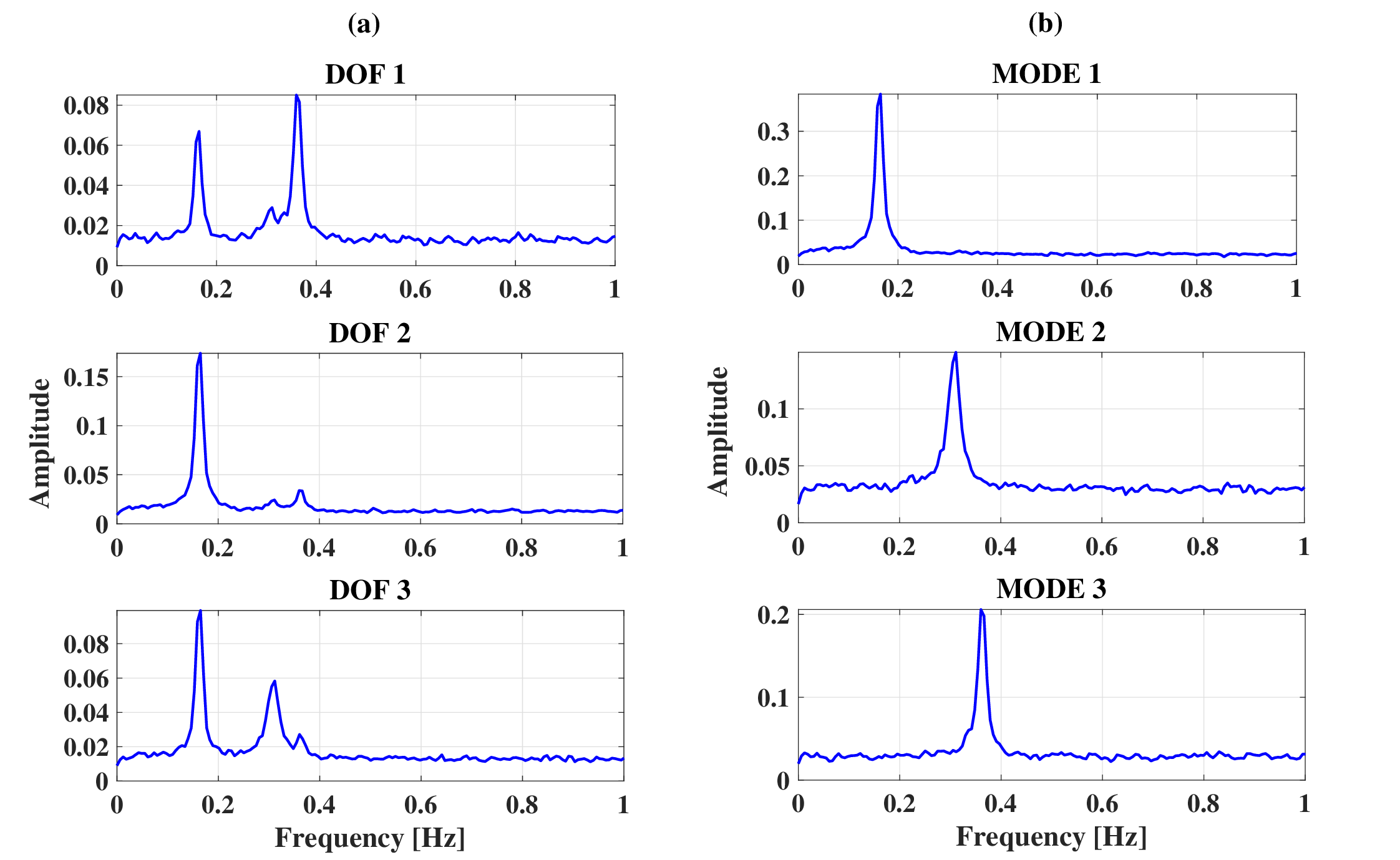}
		\caption{Frequency domain response of the (a) physical (b) modal displacement of 3DOF linear system with closely spaced modes}
		\label{3DOFCS2}
	\end{figure}
	\subsection{3DOF linear structural system with closely spaced modes}
	A 3-DOF linear vibrating structure modelled as mass, spring and dash-pot system as presented in Fig. \ref{3DOF} with its governing expression given as in Eq. (\ref{basic}). The purpose of using WGN is to excite all the modes of a structure in order to obtain accurate modal parameters and mode shapes. The Monte Carlo method is a variance reduction technique that can be used to obtain an ensemble average of the mode shape matrix, source signal, and modal parameters by generating multiple realizations of the input excitation and system response. This ensures that the results are more robust and accurate, as the average of multiple simulations helps to reduce the impact of random fluctuations and measurement noise. The state-space representation for the system subjected to a force vector $F(t)$ can be written as,
	\begin{equation}\label{statespace}
		\begin{array}{l}
			\dot U = {\bf{A}}U + {\bf{B}}F\\
			Y = {\bf{C}}U
		\end{array}
	\end{equation}
	where, $U$ = vector of states and $Y$ = system response vector governed by the $\textbf{C}$ matrix. The state matrix, $\textbf{A}$, and the excitation matrix $\textbf{B}$ are given by,
	\begin{equation}\label{system_matrix_5dof}
		\begin{array}{l}
			{\bf{A}} = \left[ {\begin{array}{*{20}{c}}
					{{{\left[ {\bf{0}} \right]}_{3 \times 3}}}&{{{\left[ {\bf{I}} \right]}_{3 \times 3}}}\\
					{ - {{\bf{M}}^{ - 1}}{\bf{K}}}&{ - {{\bf{M}}^{ - 1}}{\bf{C}}}
			\end{array}} \right]\\
			{\bf{B}} = {\left[ {\begin{array}{*{20}{c}}
						0&0&0& diag\left({\bf{M}}^{-1}\right)^T
				\end{array}} \right]^T}
		\end{array}
	\end{equation}
	where, ${\bf{M}}$,${\bf{C}}$ and ${\bf{K}}$ are the general system matrix given as,
	\begin{equation}\label{mat}
		{\bf M}  = \left[ {\begin{array}{*{20}c}
				{1.5} & {0} & {0}  \\
				{0 } & {2} & {0}  \\
				{0 } & {0} & {1.3}
		\end{array}}\right], \;
		{\bf K}  = \left[ {\begin{array}{*{20}c}
				{7} & {-2} & {0}  \\
				{-2} & {4} & {-2}  \\
				{0} & {-2} & {5}
		\end{array}} \right], \quad
		{\bf C} = \alpha {\bf M} + \beta {\bf K}
	\end{equation}
	The natural frequencies corresponding to the undertaken system parameters are $0.1649 Hz$, $0.3248 Hz$, $0.3655 Hz$. The damping ratio for the system is kept at $\zeta  = 2.0\%$ critical for each mode. The real valued mode shape can be obtained from the eigenvalue decomposition of ${\bf M^{-1} K}$ matrix, however, the actual nature of the mode shapes (real or complex) can be depicted by the eigen decomposition of the state matrix ${\bf A}$. For the sufficiently close mode shapes, the mode shape matrix will be complex in nature and for the undertaken system the real valued and complex mode shapes without normalization are evaluated as,
	\begin{equation}\label{MSCS}
		\begin{array}{ll}
			{\bf {{\phi }^{act}}}  &= \left[ {\begin{array}{*{20}c}
					{-0.2299} & {-0.4061} & {-0.6700}  \\
					{-0.6193} & {-0.1528} & {0.3051}  \\
					{-0.3437} & {0.7369} & {-0.3288}
			\end{array}} \right], \\
			{\bf \phi _{c}^{act}}  &= \left[ {\begin{array}{*{20}c}
					{0.0043 + 0.2143i} & {0.0052 + 0.2089i} & {-0.0033 - 0.3318i}  \\
					{0.0116 + 0.5774i} & {0.0020 + 0.0786i} & {0.0015 + 0.1511i}  \\
					{0.0064 + 0.3204i} & {-0.0095 - 0.3790i} & {-0.0016 - 0.1628i}
			\end{array}} \right]
		\end{array}
	\end{equation}
	\begin{figure}[htbp]
		\centering
		\includegraphics[width=0.75\textwidth]{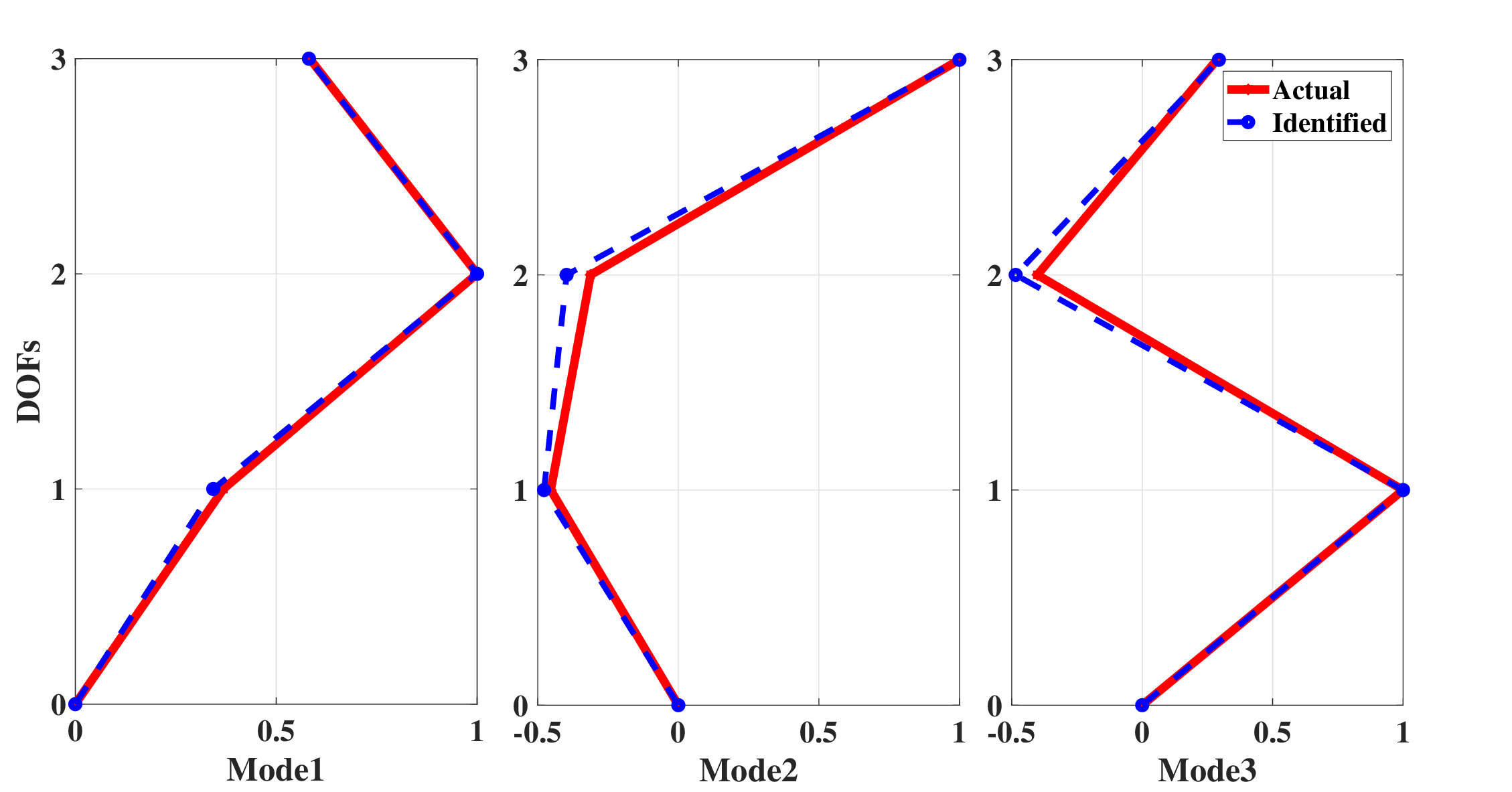}
		\caption{Actual and Identified real valued mode shapes of the 3DOF linear system with closely spaced modes}
		\label{3DOFCS3}
	\end{figure}
	The proposed framework takes the complex response (obtained through appending the 90$^\circ$ phase shifted response) as input and updates the data eigenspace recursively through FOEP to obtain the whitening matrix. Thereafter, the whitened data is utilized for the updation of initial lagged covariance matrix, which are then jointly diagonalized for the estimation of unitary matrix. Thereafter both recursive whitening and unitary matrices are used to obtain the mixing and demixing matrices which are then made use of to extract the system's source signal. Fig. \ref{3DOFCS2} describes the physical and modal displacement response in frequency domain. In Fig. \ref{3DOFCS2}(b) it can be observed that the algorithm accurately separates the modal responses of all the modes. The above observations can further be explored broadly through the MAC values, which provides a correlation between the actual and identified mode shapes. The MAC value for the identified modes are obtained as $0.9999$, $0.9910$ and $0.9835$ for modes 1,2 and 3, respectively, which validates the accuracy of the identified modes of the system. The plot of actual and identified real valued modes in Fig. \ref{3DOFCS3} demonstrates the usefulness of the proposed framework in real-world applications where traditional modal identification methods may not be effective due to the presence of complex modes.
	\subsection{3DOF linear structural system with non-proportional damping}
	As shown in the previous section, proportional damping is often used as a simplified approach to model the effect of damping in linear vibrational mechanical systems. However, there are cases in which a general viscous damping is needed to simulate the dynamic of the system with sufficient accuracy. To validate our technique for the complex modes arising due to the non-proportional damping, consider the generic differential equation given as in Eq. (\ref{basic}) with system matrices as follows: 
	\begin{equation}\label{mat}
		{\bf M}  = \left[ {\begin{array}{*{20}c}
				{3} & {0} & {0} \\
				{0 } & {2} & {0} \\
				{0 } & {0} & {1}
		\end{array}}\right], \;
		{\bf C}  = \left[ {\begin{array}{*{20}c}
				{0.1856} & {0.2290} & {-0.9702}  \\
				{0.2290} & {0.0308} & {-0.0297}  \\
				{-0.9702} & {-0.0297} & {0.1241}
		\end{array}} \right], \;
		{\bf K}  = \left[ {\begin{array}{*{20}c}
				{4} & {-2} & {0}  \\
				{-2} & {4} & {-2}  \\
				{0} & {-2} & {10}
		\end{array}} \right]
	\end{equation}	
	The input excitation is modelled as \textit{White Gaussian noise} (\textbf{WGN}) with sufficient intensity to excite all the modes properly, and then a variance reduction technique viz, Monte Carlo is employed to obtain the ensemble average of the mode shape matrix, source signal and modal parameters. The actual real valued and complex mode shapes for the undertaken system without normalization is provided in Eq. (\ref{MSNPD}).
	\begin{figure}[htbp]
		\centering
		\includegraphics[width=\textwidth]{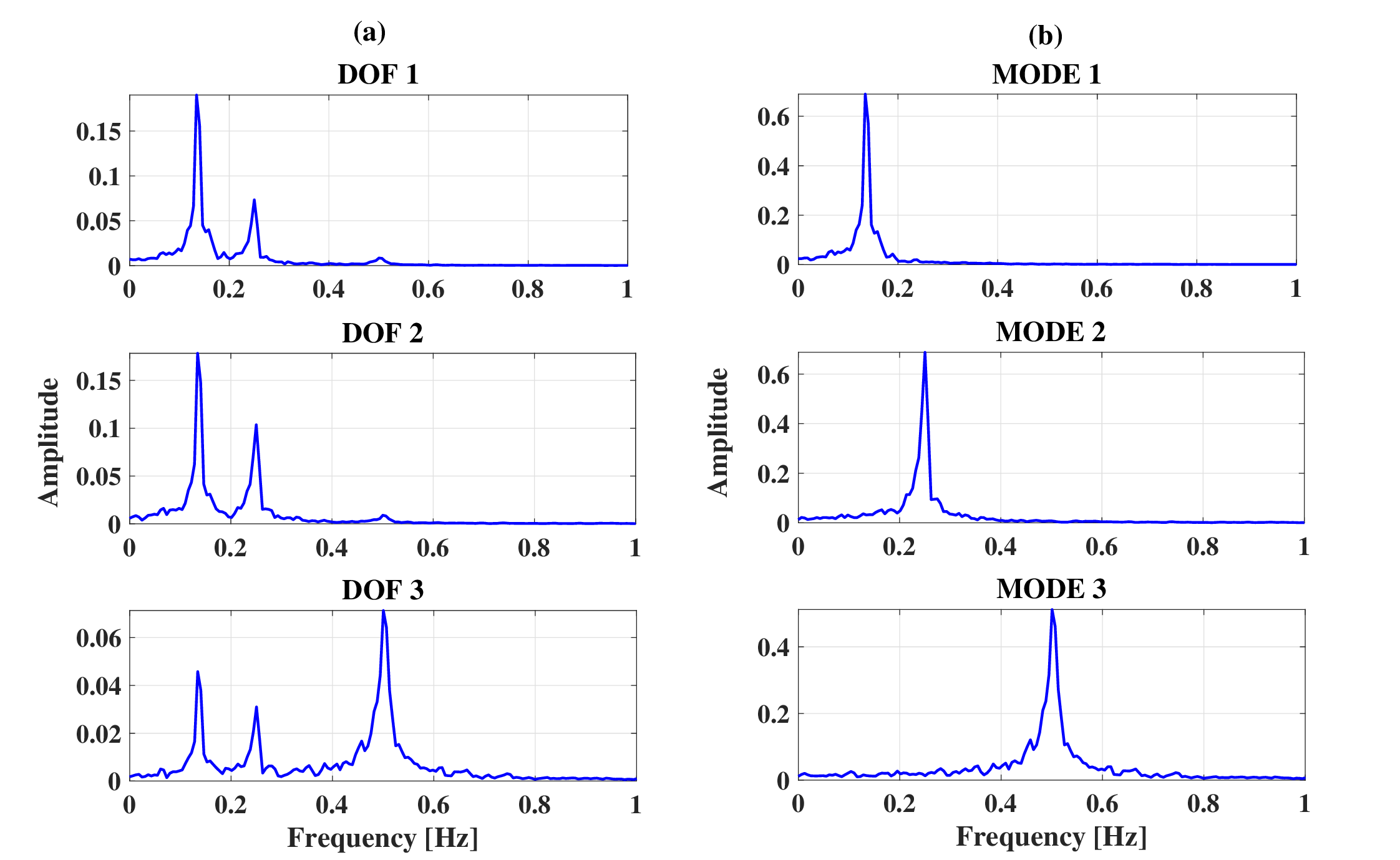}
		\caption{Frequency domain response of the (a) physical (b) modal displacement of 3DOF linear system with non-proportional damping}
		\label{3DOFNPD2}
	\end{figure}
	\begin{equation}\label{MSNPD}
		\begin{array}{ll}
			{\bf {{\phi }^{act}}}  &= \left[ {\begin{array}{*{20}c}
					0.4575  &  0.3521  &  0.0090 \\
					0.4264  & -0.5510  & -0.1206 \\
					0.0918  & -0.1446  &  0.9852
			\end{array}} \right], \\
			{\bf \phi _{c}^{act}}  &= \left[ {\begin{array}{*{20}c}
					-0.0370 - 0.0021i & -0.0115 + 0.2947i & -0.5458 + 0.0000i \\
					0.0050 + 0.0364i & -0.0059 - 0.4345i & -0.5156 - 0.0126i \\
					-0.0051 - 0.2988i & -0.0594 - 0.1167i & -0.1100 - 0.0521i
			\end{array}} \right]
		\end{array}
	\end{equation}
	\begin{figure}[htbp]
		\centering
		\includegraphics[width=0.8\textwidth]{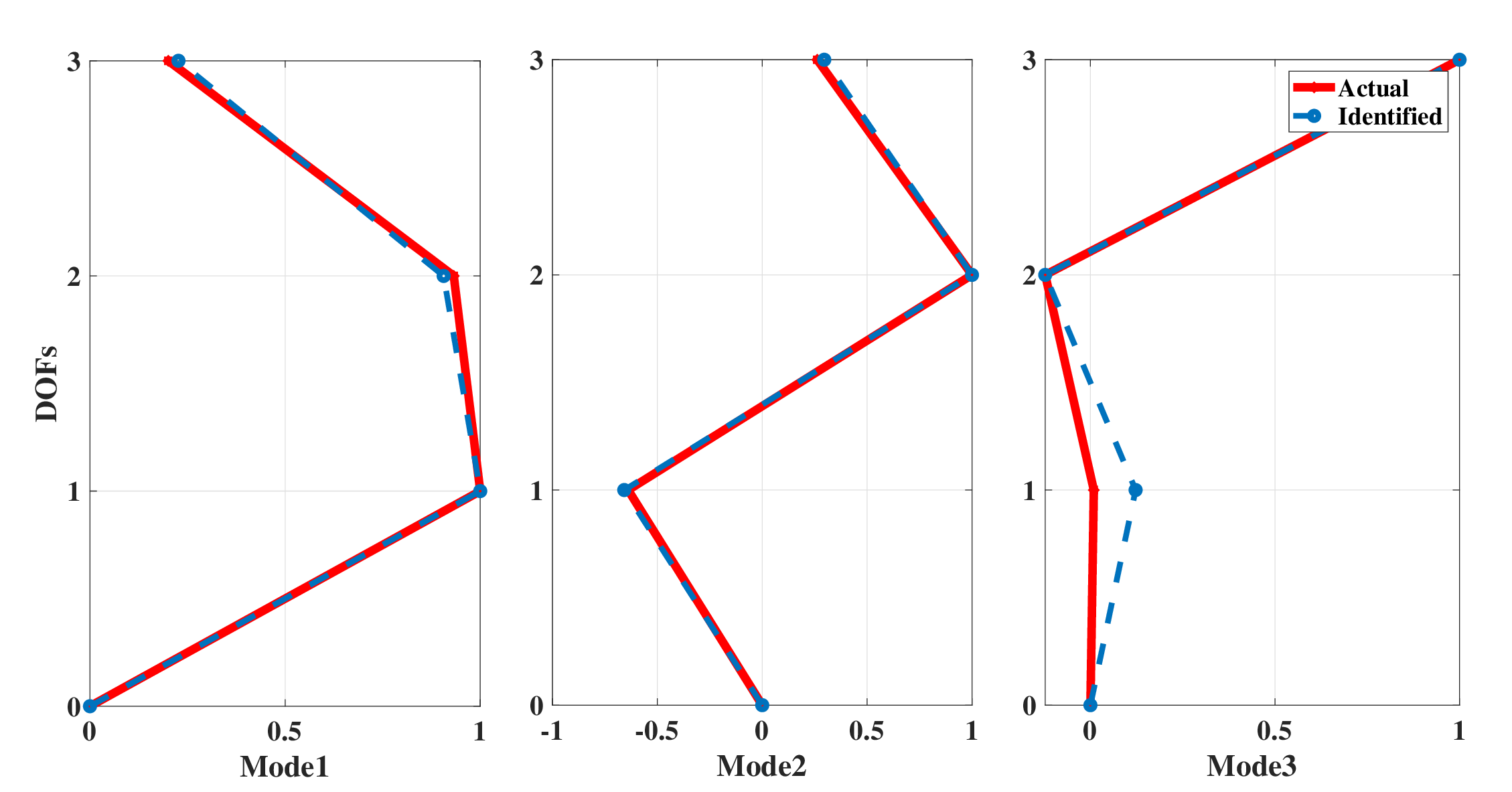}
		\caption{Actual and Identified real valued mode shapes of the 3DOF linear system with non-proportional damping}
		\label{3DOFNPD3}
	\end{figure}
	\begin{figure}[htbp]
		\centering
		\includegraphics[width=\textwidth]{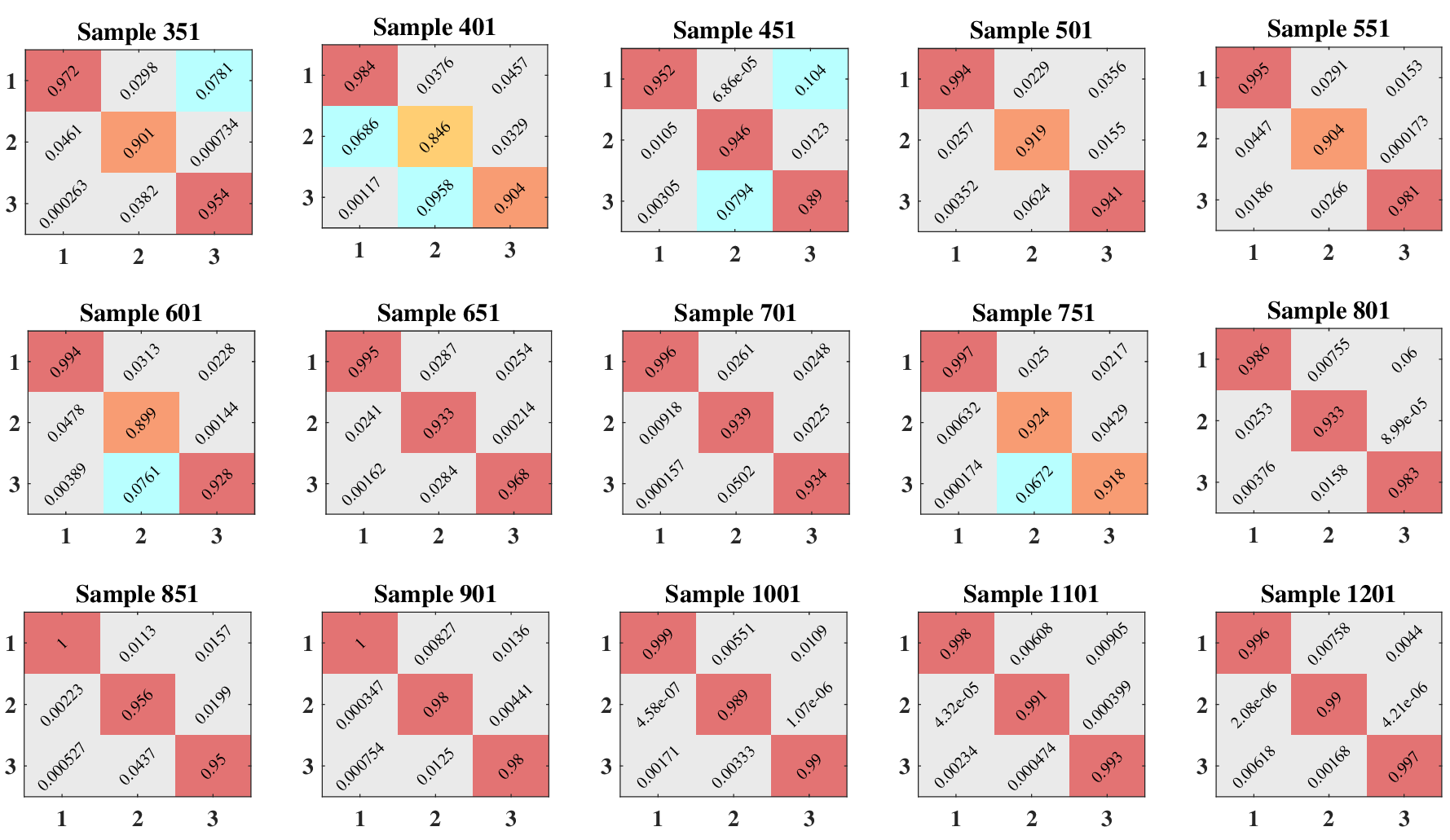}
		\caption{MAC between actual and identified modes with increase in samples representing accurate identification of modes for 3DOF system}
		\label{3DOFNPD4}
	\end{figure}
	Comparison of the estimated modal frequency response in  Fig. \ref{3DOFNPD2}(b) with the original frequency response in  Fig. \ref{3DOFNPD2}(a) provides a clear depiction of accurate modal responses separation. This is achieved through the use of proposed framework on complex responses (obtained through appending the 90$^\circ$ phase shifted response). The above observations are substantiated through the MAC values, which for the identified modes are obtained as $0.9996$, $0.9986$ and $0.9878$ for modes 1,2 and 3, respectively. This validates the accuracy of the identified modes of the system, and Fig. \ref{3DOFNPD3} shows the approximate overlap between identified modes over true modes. From the results presented in the study it is evident that the modal identification method was able to estimate the complex modes. This establishes the applicability of the technique to estimate complex modes for closely spaced modes and non-proportional damping. 
	\subsection{3DOF linear structural system with sudden change in damping}
	Most real-world phenomena have to accommodate dynamic changes in system matrices due to various operational and environmental conditions. To overcome this limitation, the proposed algorithm uses a recursive method to update the system matrices in real-time, which allows the identification of the changing modes. This real-time monitoring and updating of the modes provides more accurate information about the system behavior, which can be useful in many applications, such as real-time SHM, damage detection, and control systems. The proposed algorithm provides a robust solution for the identification of complex modes in dynamic systems, which can handle non-proportional damping, non-linearity, gyroscopic effects, and closely spaced modes. In order to validate the applicability of the proposed algorithm, a general 3DOF linear system with proportional damping is considered with the system matrices defined as,
	\begin{equation}\label{mat1}
		{\bf M}  = \left[ {\begin{array}{*{20}c}
				{3} & {0} & {0}  \\
				{0 } & {2} & {0}  \\
				{0 } & {0} & {2}
		\end{array}}\right], \;
		{\bf K}  = \left[ {\begin{array}{*{20}c}
				{4} & {-2} & {0}  \\
				{-2} & {4} & {-2}  \\
				{0} & {-2} & {10}
		\end{array}} \right], \quad
		{\bf C} = \alpha {\bf M} + \beta {\bf K}
	\end{equation}
	Here, the damping coefficient is 2$\%$ and the system is excited with \textbf{WGN} for a time period of 50s and sampling frequency is taken as 50Hz. This system is considered to be in its prime state and has real valued modes. At 25s the damping matrix suddenly becomes non-proportional and is given as,
	\begin{equation}\label{dampmat}
		{\bf C}  = \left[ {\begin{array}{*{20}c}
				{0.1856} & {0.2290} & {-0.9702}  \\
				{0.2290} & {0.0308} & {-0.0297}  \\
				{-0.9702} & {-0.0297} & {0.1241}
		\end{array}} \right]
	\end{equation}
	The perturbations in the damping matrix convert the real-valued modes into complex-valued modes -- presenting a more formidable challenge to conventional blind source separation (BSS) algorithms -- as the precision of the mode identifications can not be substantiated in practical conditions, nor can the alterations be monitored. Nonetheless, the proposed algorithm encompasses the requisite tools to effectively monitor such alterations and accurately determine both real-valued modes (i.e. proportional damping within the interval of 0 to 25 seconds) and complex-valued modes (i.e. non-proportional damping within the interval of 25 to 50 seconds). To further verify the applicability of the proposed algorithm in practical scenarios, a benchmark system is described in the subsequent section.
	\begin{figure}
		\centering
		\includegraphics[width=0.7\textwidth]{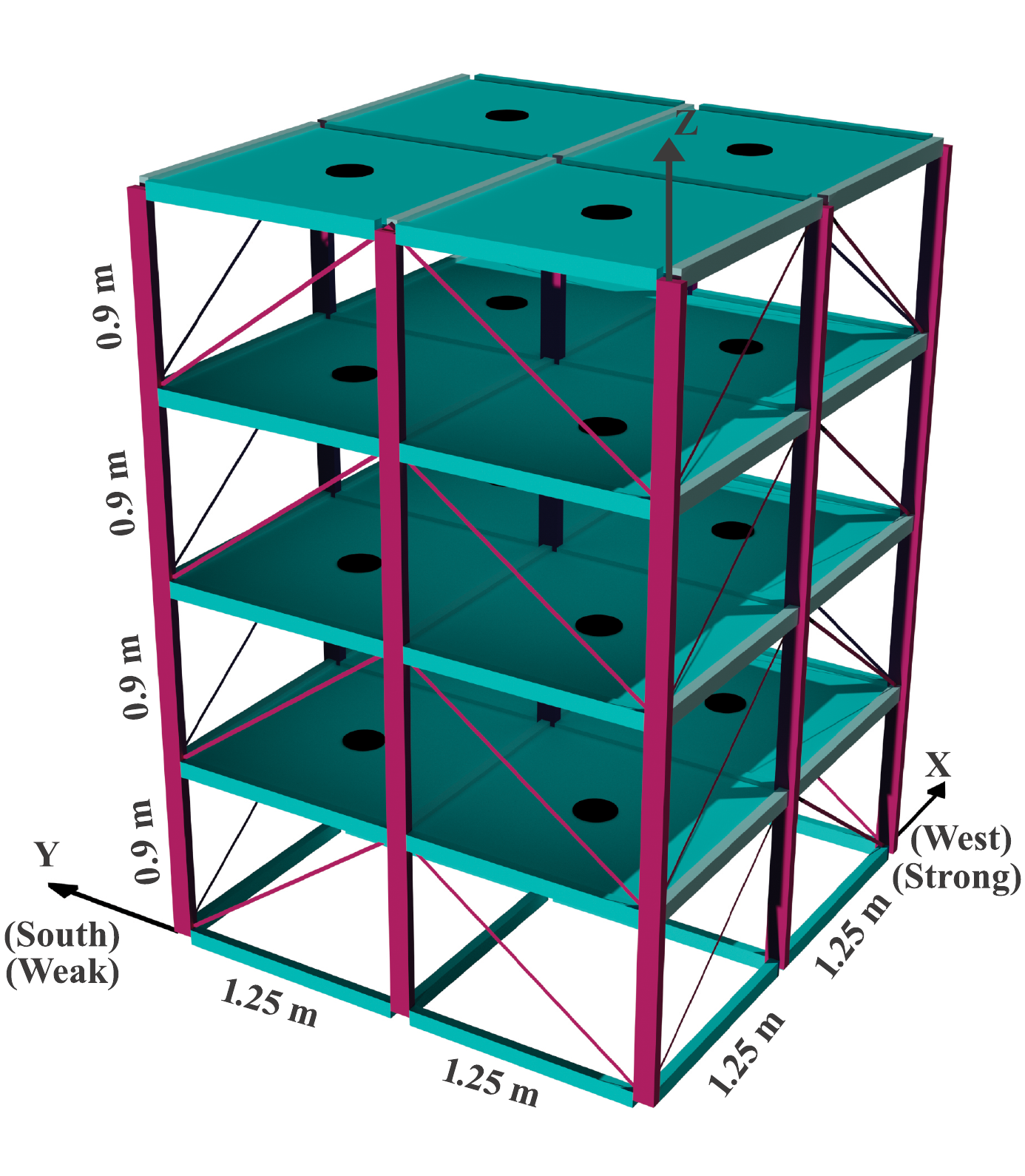}
		\caption{ASCE-SHM benchmark model adopted from \cite{johnson2004phase}}
		\label{12DOF1}
	\end{figure}
	\subsection{IASC-ASCE SHM benchmark structure with complex modes}
	Having established the viability of the proposed algorithm through numerical simulation of simple systems, this study progresses to the examination of its application to the IASC-ASCE Structural Health Monitoring (SHM) Benchmark structure. The latter constitutes a quarter-scale model of a four-story, two-bay by two-bay steel frame structure located at the University of British Columbia's Earthquake Engineering Research Laboratory. It boasts a planar dimension of 2.5 m x 2.5 m and a height of 3.6 m, with members fabricated from hot rolled grade 300 W steel, possessing a nominal yield stress of 300 MPa. The structural sectional properties can be consulted in \cite{johnson2004phase}. The parameters utilized to generate the response of the benchmark structure are: a damping ratio of 0.01, a sampling time step size of 0.002s, a total time duration of 50s, a noise level of 1, and a force coefficient of 150.
	\begin{figure}[htbp]
		\centering
		\includegraphics[width=0.9\textwidth]{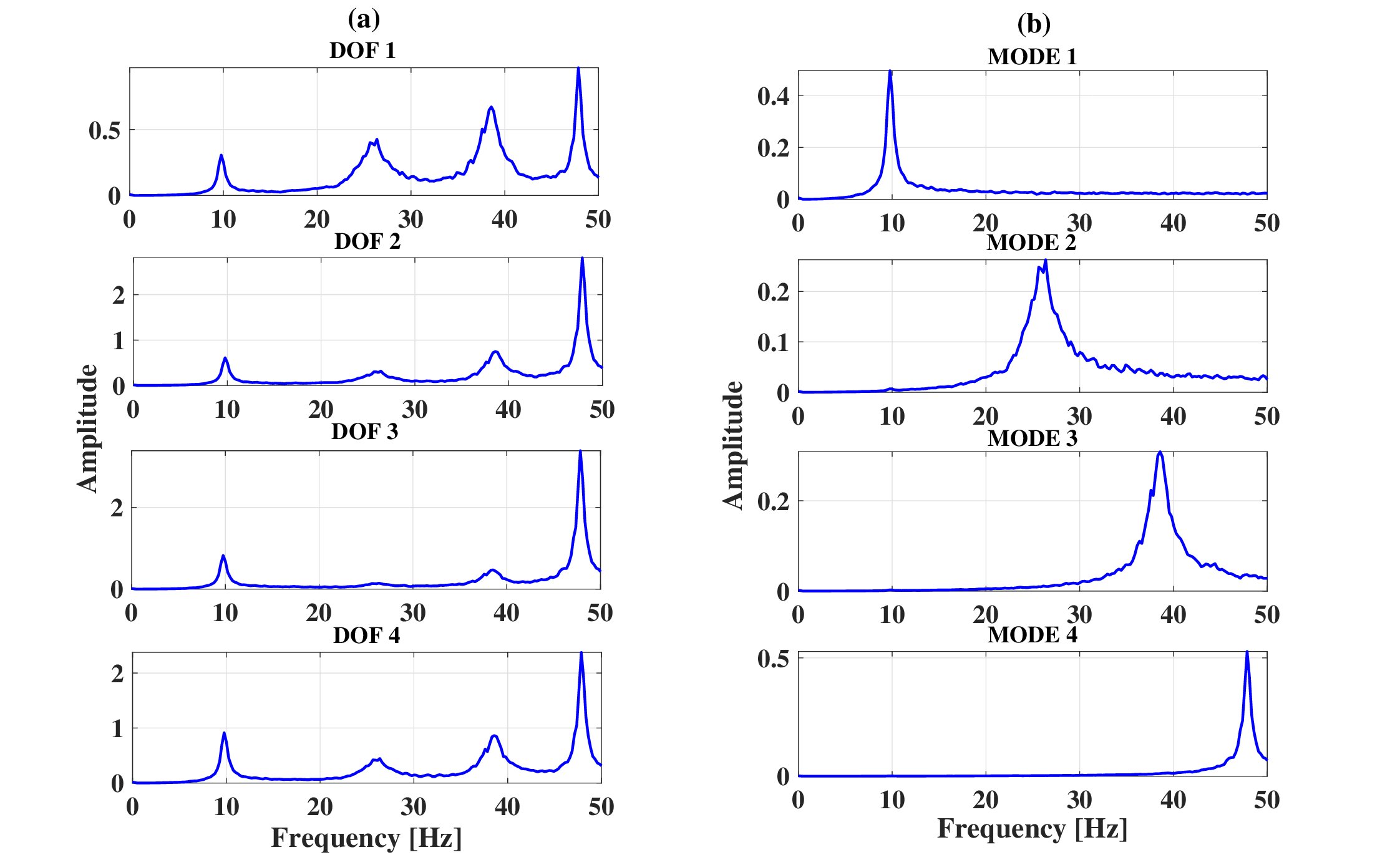}
		\caption{Frequency domain response of the (a) physical (b) modal displacement of modified ASCE SHM benchmark structure}
		\label{3DOFBEN2}
	\end{figure}
	\begin{figure}[htbp]
		\centering
		\includegraphics[width=0.9\textwidth]{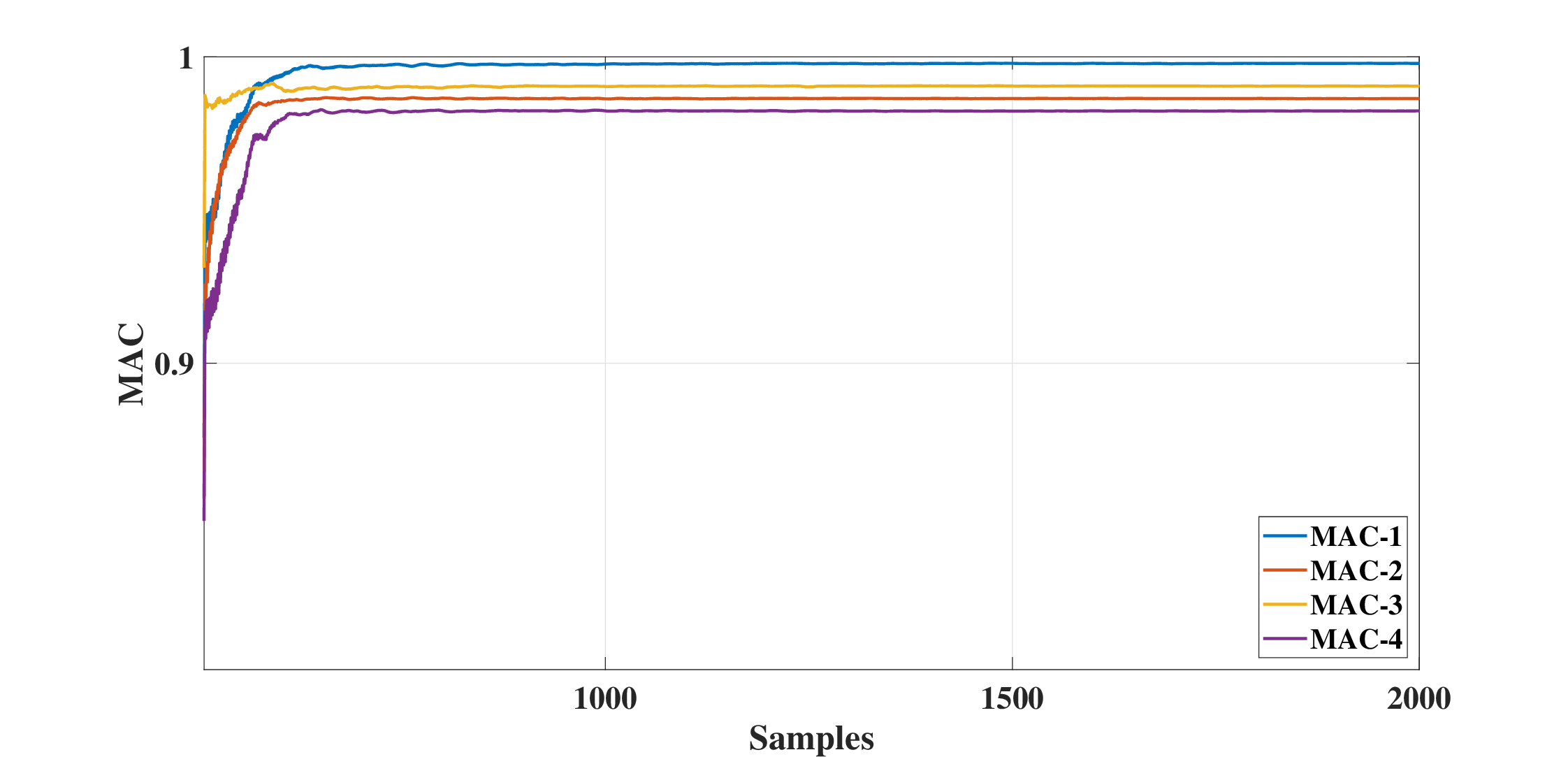}
		\caption{MAC between actual and identified modes representing accurate identification of modes for ASCE-SHM benchmark structure}
		\label{3DOFBEN3}
	\end{figure}
	\begin{figure}[htbp]
		\centering
		\includegraphics[width=0.8\textwidth]{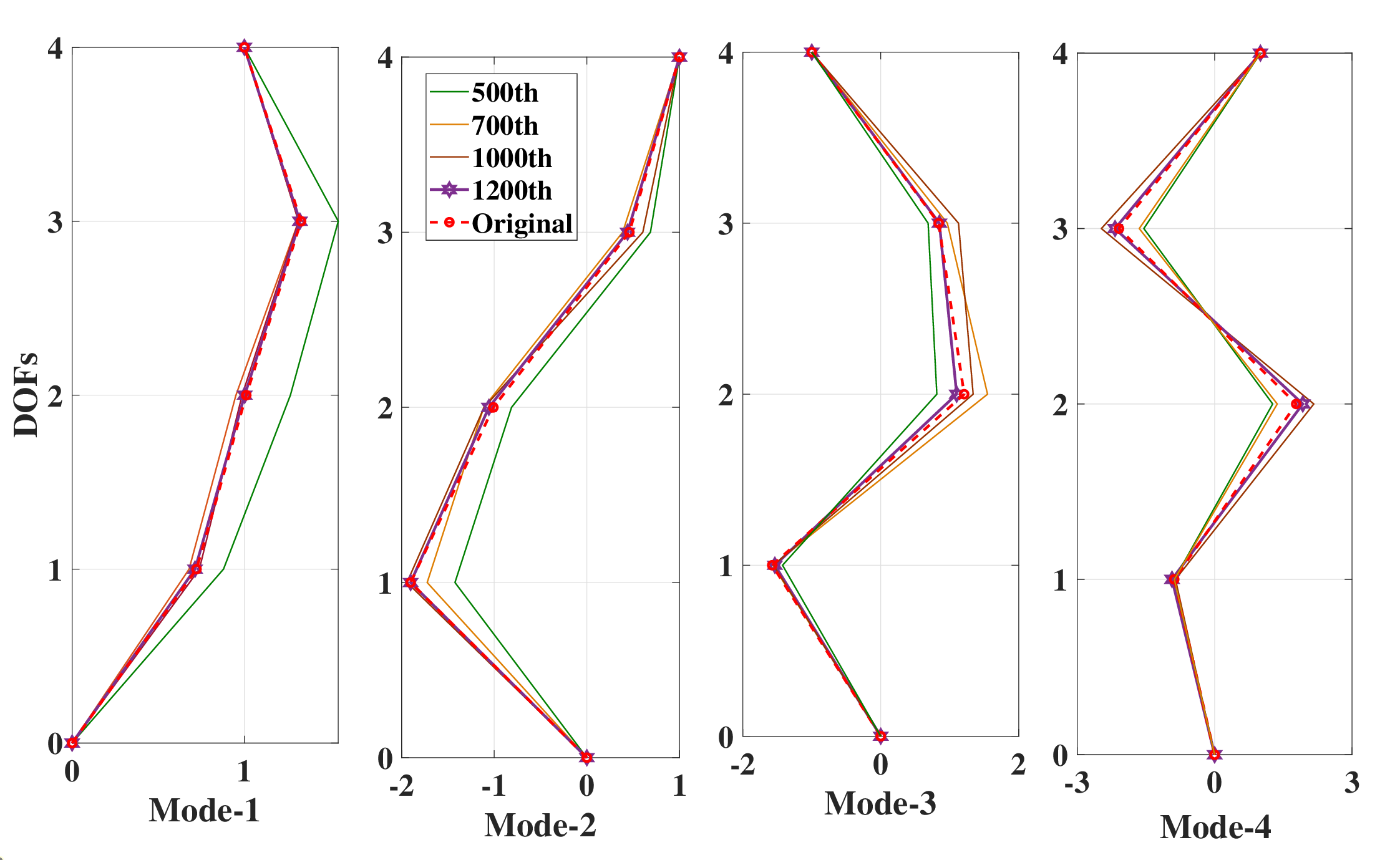}
		\caption{Original and identified real valued mode shapes of the ASCE SHM benchmark system with non-proportional damping for different samples}
		\label{3DOFBEN4}
	\end{figure}
	In the IASC-ASCE SHM Benchmark structure, the columns are aligned in the x direction, which exhibits superior resistance to bending, while the floor beams are aligned to withstand strong vertical bending. Each bay on each floor features a single floor slab, with the first, second, third, and fourth levels comprising four slabs weighing 800 kg, 600 kg, 600 kg, and 400 kg, respectively, for the symmetrical building configuration. In contrast, the fourth level in the unsymmetrical building configuration includes three 400 kg slabs and one 550 kg slab. To generate simulated response data, a finite element model is devised for various conditions. In the absence of damage, a 12-degree-of-freedom (DOF) shear-building model that restricts all motion (with the exception of translations in the x and y directions and a single rotation $\theta$ about the center column) by modeling the floor beams and floor slabs as rigid bodies is created, resulting in three DOFs per floor. The columns and floor beams are modeled as Euler-Bernoulli beams in the finite element model. A diagram of the analytical model is shown in Fig. \ref{12DOF1} (case 1) with no damage in the structure (damage pattern 0) is taken for the validation of the proposed algorithm. For the undamaged case, the mass (${\bf{M}}$) and stiffness (${\bf{K}}$) matrices of the structure are given by the following matrices:
	\begin{equation}
		\resizebox{\textwidth}{!}{$
			\begin{array}{l}
				{\bf{M}} = \left[ {\begin{array}{*{20}{c}}
						{3452.4}&0&0&0&0&0&0&0&0&0&0&0\\
						0&{3452.4}&0&0&0&0&0&0&0&0&0&0\\
						0&0&{3819.4}&0&0&0&0&0&0&0&0&0\\
						0&0&0&{2652.4}&0&0&0&0&0&0&0&0\\
						0&0&0&0&{2986.1}&0&0&0&0&0&0&0\\
						0&0&0&0&0&{2652.4}&0&0&0&0&0&0\\
						0&0&0&0&0&0&{2652.4}&0&0&0&0&0\\
						0&0&0&0&0&0&0&{2986.1}&0&0&0&0\\
						0&0&0&0&0&0&0&0&{1809.9}&0&0&0\\
						0&0&0&0&0&0&0&0&0&{1809.9}&0&0\\
						0&0&0&0&0&0&0&0&0&0&{1809.9}&0\\
						0&0&0&0&0&0&0&0&0&0&0&{2056.9}
				\end{array}} \right]\text{kg}\\
			\end{array}$}
	\end{equation}
	\begin{equation}
		\resizebox{\textwidth}{!}{$
			\begin{array}{l}
				{\bf{K}} = \left[ {\begin{array}{*{20}{c}}
						{213.20}&0&0&{ - 106.60}&0&0&0&0&0&0&0&0\\
						0&{135.81}&0&0&{ - 67.90}&0&0&0&0&0&0&0\\
						0&0&{464.04}&0&0&{ - 232.02}&0&0&0&0&0&0\\
						{ - 106.60}&0&0&{213.20}&0&0&{ - 106.60}&0&0&0&0&0\\
						0&{ - 67.90}&0&0&{135.81}&0&0&{ - 67.90}&0&0&0&0\\
						0&0&{ - 232.02}&0&0&{464.04}&0&0&{ - 232.02}&0&0&0\\
						0&0&0&{ - 106.60}&0&0&{213.20}&0&0&{ - 106.60}&0&0\\
						0&0&0&0&{ - 67.90}&0&0&{135.81}&0&0&{ - 67.90}&0\\
						0&0&0&0&0&{ - 232.02}&0&0&{464.04}&0&0&{ - 232.02}\\
						0&0&0&0&0&0&{ - 106.60}&0&0&{106.60}&0&0\\
						0&0&0&0&0&0&0&{ - 67.90}&0&0&{67.90}&0\\
						0&0&0&0&0&0&0&0&{ - 232.02}&0&0&{232.02}
				\end{array}} \right]\text{MN/m}
			\end{array}$}
	\end{equation}
	The IASC-ASCE SHM Benchmark structure is symmetric, except in cases where damage results in loss of symmetry, and the applied forces are limited to the y direction. In such scenarios, the output measurements indicate no response in the x or $\theta$ direction, and only the y direction measurements are utilized in the identification process to reduce any numerical instability. Given that the system in question has sparsely separated modes and mass- and stiffness-proportional damping, the actual modes are real and can be obtained as follows:
	\begin{equation}\label{MSBENR}
		\begin{array}{ll}
			{\bf {{\phi }^{act}}}  &= \left[ {\begin{array}{*{20}c}
					0.2422  & -0.6226 &  0.5029  &  0.2129 \\
					0.4414  & -0.4297 &  -0.5032 &  -0.5591 \\
					0.5803  &  0.1956  & -0.3494  &  0.6559 \\
					0.6399 &  0.6241  &  0.6098 &  -0.4603
			\end{array}} \right]
		\end{array}
	\end{equation}
	To introduce complex modes into the benchmark structure for the purposes of our study, an additional mass with a stiffness and damper is added to the top story, affecting only the y direction. The new dynamic system has a mass of $172.6$ kg, a stiffness of $6.8$ MN/m, and a 2\% damping ratio. With the addition of this new dynamic system, the modes of the system in question become complex in the y direction and can be characterized as follows:

	\begin{equation}\label{MSBEND}
		\resizebox{\textwidth}{!}{$
			\begin{array}{ll}
				{\bf \phi _{c}^{act}}  &= \left[ {\begin{array}{*{20}c}
						0.0980 - 0.6579i &  0.2123 - 2.0306i &  0.1644 + 3.8407i & -0.2735 - 3.7886i \\
						-0.0310 + 1.8428i &  0.1897 + 2.1281i & -0.2647 + 2.7033i & -0.2480 - 7.2326i \\
						-0.0123 - 2.1900i & -0.0904 + 1.4109i & -0.2072 - 1.1655i & -0.2244 - 9.6458i \\
						0.0105 + 1.5428i & -0.0524 - 2.5213i & -0.0388 - 3.8303i & -0.2230 -10.6845i
				\end{array}} \right]
			\end{array}
			$}
	\end{equation}
	The proposed algorithm was also applied to a benchmark structure subjected to stochastic excitation. The comparison between the physical and estimated modal responses in the frequency domain is presented in Figures \ref{3DOFBEN2}(a) and \ref{3DOFBEN2}(b), respectively, which shows a clear separation of the modal responses, indicating that the modes are indeed complex. The accuracy of the identified modes was further validated through the Maximum Absolute Coefficient (MAC) values, which are plotted against the number of samples in Figure \ref{3DOFBEN3}. The comparison between the actual and identified real-valued modes, as shown in Figure \ref{3DOFBEN4}, demonstrates a close match between the identified modes and the true modes, as evidenced by the high MAC values. These results demonstrate the applicability of the proposed algorithm for accurately estimating complex modes in a wide range of numerical simulations and real-life structural models.
	
	\section{Conclusions}
	The findings of this research have firmly established the capability of the proposed algorithm in identifying complex modes in a wide range of dynamic systems in real-time. The algorithm is built upon the principle of first-order eigen perturbation (FOEP) and is designed to synchronize with the complex response, updating the second-order statistics recursively. Detection of features of
	interest may not necessarily be related to sudden or gradual damage, but can also
	be related to parameters representing their performance with respect to control,
	repair or change in other operational conditions. In many practical vibrating systems, the modes can either inherently be or become complex due to dynamic modifications of the system. This study therefore provides compelling evidence for the real-time estimation of complex modes and the accurate identification of modal properties, even in the presence of dynamic changes to the system's characteristics. The proposed algorithm effectively overcomes the limitations of traditional batch algorithms and real-time techniques, providing accurate real-time identification of complex modes even in the presence of closely spaced modes, non-proportional damping, dynamically changing damping matrices, and the addition of new dynamical systems. The results of this study have been demonstrated through extensive numerical simulations and benchmark case studies, making it a highly reliable and robust approach for real-time evaluation of complex modes.

	\section*{Acknowledgement} Satyam Panda acknowledges the financial support received from Prime Minister Research Fellowship. Budhaditya Hazra gratefully acknowledge the financial support received from Science and Engineering Research Board (SERB), Department of Science and Technology (DST), Government of India (Project no. IMP/2019/000276).

	\newcommand{\noopsort}[1]{} \newcommand{\printfirst}[2]{#1}
	\newcommand{\singleletter}[1]{#1} \newcommand{\switchargs}[2]{#2#1}

\end{document}